# Clinically Interpretable Survival Risk Stratification in Head and Neck Cancer Using Bayesian Networks and Markov Blankets

Keyur D. Shah[1*], Ibrahim Chamseddine[2], Xiaohan Yuan[3], Sibo Tian[1], Richard Qiu[1], Jun Zhou[1], Anees Dhabaan[1], Hania Al-Hallaq[1], David S. Yu[1], Harald Paganetti[2] and Xiaofeng Yang[1*]

[1]Department of Radiation Oncology, Winship Cancer Institute, Emory University, Atlanta, GA
[2]Department of Radiation Oncology, Mass General Brigham and Harvard Medical School, Boston, MA
[3]Department of Biomedical Engineering, Georgia Institute of Technology, Atlanta, GA

[*] Corresponding Authors:
Keyur D. Shah, kshah41@emory.edu,
Xiaofeng Yang, xiaofeng.yang@emory.edu

**Key Words:** Head and Neck Cancer, Survival Prediction, Machine Learning, Causal Inference, Bayesian Network

# Abstract

**Purpose:** To identify a clinically interpretable subset of survival-relevant features in head and neck (H&N) cancer using Bayesian Network (BN) and evaluate its prognostic and causal utility.

**Methods and Materials:** We used the RADCURE dataset, consisting of 3,346 patients with H&N cancer treated with definitive (chemo)radiotherapy. A probabilistic BN was constructed to model dependencies among clinical, anatomical, and treatment variables. The Markov Blanket (MB) of two-year survival (SVy2) was extracted and used to train a logistic regression model. After excluding incomplete cases, a temporal split yielded a train/test (2,174/820) dataset using 2007 as the cutoff year. Model performance was assessed using area under the receiver operating characteristic (ROC) curve (AUC), concordance index (C-index), and Kaplan-Meier (KM) survival stratification. Model fit was further evaluated using a log-likelihood ratio (LLR) test. Causal inference was performed using do-calculus interventions on MB variables.

**Results:** The MB of SVy2 included 6 clinically relevant features: Eastern Cooperative Oncology Group (ECOG) performance status, T-stage, HPV status, disease site, the primary gross tumor volume (GTVp), and treatment modality. The model achieved an AUC of 0.65 and C-index of 0.78 on the test dataset, significantly stratifying patients into high- and low-risk groups (log-rank $p < 0.01$). Model fit was further supported by a log-likelihood ratio of 70.32 ($p < 0.01$). Subgroup analyses revealed strong performance in HPV-negative (AUC = 0.69, C-index = 0.76), T4 (AUC = 0.69, C-index = 0.80), and large-GTV (AUC = 0.67, C-index = 0.75) cohorts, each showing significant KM separation. Causal analysis further supported the positive survival impact of ECOG 0, HPV-positive status, and chemoradiation.

**Conclusions:** A compact, MB-derived BN model can robustly stratify survival risk in H&N cancer. The model's structure enables explainable prognostication and supports individualized decision-making across key clinical subgroups.

# 1. Introduction

Head and neck (H&N) cancer accounts for over 900,000 new cases and approximately 400,000 deaths globally in 2020, making it the seventh most common cancer worldwide (1). These malignancies span multiple anatomical subsites, including the oropharynx, hypopharynx, larynx, and oral cavity, and exhibit substantial biological and clinical heterogeneity (2, 3). Survival outcomes are influenced by a range of factors, including tumor site and stage, HPV status, patient performance status, comorbidities, and treatment modality (4, 5). Despite progress in radiation and systemic therapies, accurately predicting individual survival remains a major challenge (6, 7). Prior studies have investigated associations between clinical variables and survival in patients with H&N cancer, identifying prognostic factors such as T-stage, HPV status, tumor site, and Eastern Cooperative Oncology Group (ECOG) performance status (2, 8). Although informative, these studies often examine predictors in isolation, without modeling their interdependencies. However, such approaches may fail to capture the underlying structure or interplay between variables that ultimately drive survival. Understanding these interdependencies is essential for improving risk stratification and moving toward more personalized treatment strategies.

Bayesian Networks (BNs) offer a probabilistic, graphical framework that models conditional dependencies between variables and facilitates both prediction and interpretability (9–11). Unlike many traditional statistical approaches, BNs allow for visual mapping of relationships among clinical features and outcomes, enabling clinicians to explore potential pathways and variable interactions. A particularly valuable property of BNs is the ability to identify the Markov Blanket (MB) of an outcome variable—a minimal set of predictors that renders the outcome conditionally independent of all other variables in the model (12). This concept offers an elegant balance between model simplicity and predictive power, while maintaining clinical interpretability.

In this study, we apply BN modeling to a large, curated cohort of patients with H&N cancer. We aim to identify a minimal, clinically relevant set of features for survival prediction, evaluate its performance against full-variable models, and assess the model's ability to stratify risk in key clinical subgroups. Our approach emphasizes interpretability and reusability, offering a foundation for transparent decision support and future causal inference applications.

## 2. Materials and Methods

### 2.1 Data set

This study utilized clinical and imaging data from the RADCURE dataset (13), which includes 3,346 patients with H&N cancer treated with definitive radiotherapy or chemoradiotherapy. All patients included in the analysis had complete clinical information and available segmentations for the primary gross tumor volume (GTVp). Exclusion criteria included missing key clinical variables or unreliable GTVp volume extraction.

The primary outcome for this study was 2-year overall survival (SVy2), defined as survival of at least 24 months from the start of radiotherapy. Survival status was determined by comparing radiotherapy start dates with the last follow-up or date of death. Patients who survived beyond 24 months were labeled as SVy2 = 1, while those with survival less than 24 months were labeled as SVy2 = 0. Patients with less than 24 months of follow-up and no recorded date of death were conservatively labeled as SVy2 = 0.

We split the dataset into training and test subsets based on year of treatment, following the stratification scheme used in the RADCURE Challenge (14). Specifically, patients whose radiotherapy start year was 2007 or earlier were assigned to the training cohort, and those treated after 2007 were assigned to the test cohort. While these years are derived from de-identified date offsets for de-identification, the stratification mimics a real-world temporal split representative of treatment years ranging approximately from 2005 to 2017.

Clinical features were standardized through a structured preprocessing pipeline. ECOG performance status values were converted to numeric scores and imputed using the cohort median where missing. Tumor staging information (T, N, M) was harmonized according to American Joint Committee on Cancer (AJCC) guidelines and mapped to ordinal categories. Histopathology was categorized into common diagnostic groups, including squamous cell carcinoma, adenocarcinoma, neuroendocrine tumors, sarcomas, melanomas, and a residual "other" category. HPV status, patient sex, treatment modality, and primary disease site were also cleaned and encoded. Disease sites were grouped into major H&N subsites—including the oropharynx, larynx, nasopharynx, oral cavity, and hypopharynx—with less common locations grouped into "Other H&N" or "Unknown." Smoking history was quantified using pack-years, cleaned for consistency, and imputed using the median.

In addition to clinical features, we incorporated a spatial anatomical variable that reflects the proximity of the primary tumor to major vasculature. Blood vessels were segmented using the TotalSegmentator deep learning model applied to the planning CT images (15). The carotid arteries, subclavian arteries and brachiocephalic trunk were merged into a single binary mask. The 3D Euclidean distance from the centroid of each patient's GTVp to the nearest point on this combined vessel mask was computed and included as a continuous variable in the analysis (GTV-to-vessel distance).

A complete list of included variables, feature encoding schemes, and value distributions is summarized in Table 1. The full modeling pipeline is summarized in the workflow shown in Figure 1.

## 2.2 Patient Characteristics

A total of 2,994 patients were included in the analysis, with 2,174 patients assigned to the training dataset and 820 patients assigned to the temporally separated test dataset. Statistical comparisons between training and test cohorts were conducted using Mann-Whitney U tests for continuous variables and chi-square tests for categorical variables. Table 1 summarizes key demographic, clinical, and treatment characteristics across the two cohorts. The mean age was slightly higher in the test cohort (63.7) compared to the training cohort (62.3), with the difference reaching statistical significance ($p < 0.01$). The sex distribution was balanced between cohorts, with approximately 80% of patients identified as male in both sets.

Other features demonstrated significant distributional shifts between training and test sets. ECOG performance status was lower in the training set, with 62.2% of patients having an ECOG of 0 compared to 54.0% in the test cohort ($p < 0.01$). HPV status also differed significantly ($p < 0.01$), with a higher proportion of HPV-positive cases in the training set. Similarly, patients in the training set had slightly higher cumulative smoking exposure ($p < 0.01$). The distribution of disease sites, T/N staging, histopathology, and treatment modalities was largely consistent across cohorts. As expected, given the temporal split, the outcome distribution differed between cohorts: 77.7% of patients in the training cohort survived at least two years post-treatment ($SVy2 = 1$), compared to 52.4% in the test cohort (p < 0.01). These differences are expected with evolving clinical practices, demographics shifts, and longer follow-up times in earlier cohorts.

**Table 1.** Baseline demographic and clinical characteristics of the training and testing cohorts. Continuous variables are presented as mean ± standard deviation; categorical variables are reported as counts with percentages. P-values reflect comparisons between training and testing sets using Mann-Whitney U tests for continuous variables and chi-square tests for categorical variables.

| Feature | Training | Testing | p-value |
|---|---|---|---|
| Age (years) | 62.3 ± 11.8 | 63.7 ± 11.3 | <0.01 |
| **Sex** | | | |
| Male | 1725 (79.3%) | 658 (80.2%) | 0.62 |
| Female | 449 (20.7%) | 162 (19.8%) | |
| **ECOG PS** | | | |
| 0 | 1352 (62.2%) | 443 (54.0%) | <0.01 |
| 1 | 611 (28.1%) | 346 (42.2%) | |
| 2 | 168 (7.7%) | 28 (3.4%) | |
| 3 | 36 (1.7%) | 3 (0.4%) | |
| 4 | 7 (0.3%) | 0 (0.0%) | |
| Smoking Packs per year | 26.3 ± 25.0 | 23.0 ± 21.9 | <0.01 |
| **Disease Site** | | | |
| Esophagus | 23 (1.1%) | 10 (1.2%) | 0.08 |
| Hypopharynx | 124 (5.7%) | 33 (4.0%) | |
| Larynx | 635 (29.2%) | 223 (27.2%) | |
| Nasopharynx | 249 (11.5%) | 101 (12.3%) | |
| Oropharynx | 964 (44.3%) | 377 (46.0%) | |
| Other H&N | 53 (2.4%) | 34 (4.1%) | |
| Unknown | 126 (5.8%) | 42 (5.1%) | |
| **T** | | | |
| 0 | 21 (1.0%) | 8 (1.0%) | 0.92 |
| 1 | 440 (20.2%) | 171 (20.9%) | |
| 2 | 619 (28.5%) | 246 (30.0%) | |
| 3 | 626 (28.8%) | 223 (27.2%) | |
| 4 | 429 (19.7%) | 156 (19.0%) | |
| Unknown | 39 (3.8%) | 16 (2.0%) | |
| **N** | | | |
| 0 | 815 (37.5%) | 282 (34.4%) | 0.14 |
| 1 | 210 (9.7%) | 98 (12.0%) | |
| 2 | 1018 (46.8%) | 400 (48.8%) | |
| 3 | 122 (5.6%) | 36 (4.4%) | |
| Unknown | 9 (0.4%) | 4 (0.5%) | |
| **M** | | | |
| 0 | 2162 (99.4%) | 814 (99.3%) | 0.07 |
| 1 | 0 (0.0%) | 2 (0.2%) | |
| Unknown | 12 (0.6%) | 4 (0.5%) | |
| **Stage** | | | |
| I | 284 (13.1%) | 90 (11.0%) | 0.24 |
| II | 273 (12.6%) | 110 (13.4%) | |
| III | 416 (19.1%) | 154 (18.8%) | |
| IV | 1186 (54.6%) | 455 (55.5%) | |
| Unknown | 15 (0.7%) | 11 (1.3%) | |
| **Pathology** | | | 0.10 |
|     Squamous | 1886 (86.8%) | 696 (84.9%) | |
|     Adeno/Basal/Sebaceous | 15 (0.7%) | 6 (0.7%) | |
|     Neuroendocrine | 10 (0.5%) | 2 (0.2%) | |
|     Sarcoma | 3 (0.1%) | 6 (0.7%) | |
|     Melanoma | 1 (0.0%) | 1 (0.1%) | |
|     Unknown | 259 (11.9%) | 109 (13.3%) | |
| **HPV Status** | | | <0.01 |
|     Negative | 344 (15.8%) | 197 (24.0%) | |
|     Positive | 637 (29.3%) | 312 (38.0%) | |
|     Unknown | 1193 (54.9%) | 311 (37.9%) | |
| **Treatment Modality** | | | <0.01 |
|     ChemoRT | 887 (40.8%) | 353 (43.0%) | |
|     Postop RT alone | 1 (0.0%) | 1 (0.1%) | |
|     RT + EGFRI | 63 (2.9%) | 7 (0.9%) | |
|     RT alone | 1223 (56.3%) | 459 (56.0%) | |
| Primary GTV Volume ($cm^3$) | 26.2 ± 32.4 | 26.7 ± 38.2 | 0.09 |
| Primary GTV to vessel distance (mm) | 51.0 ± 28.0 | 51.6 ± 31.1 | 0.58 |
| **SVy2** | | | <0.01 |
|     No (0) | 484 (22.3%) | 390 (47.6%) | |
|     Yes (1) | 1690 (77.7%) | 430 (52.4%) | |

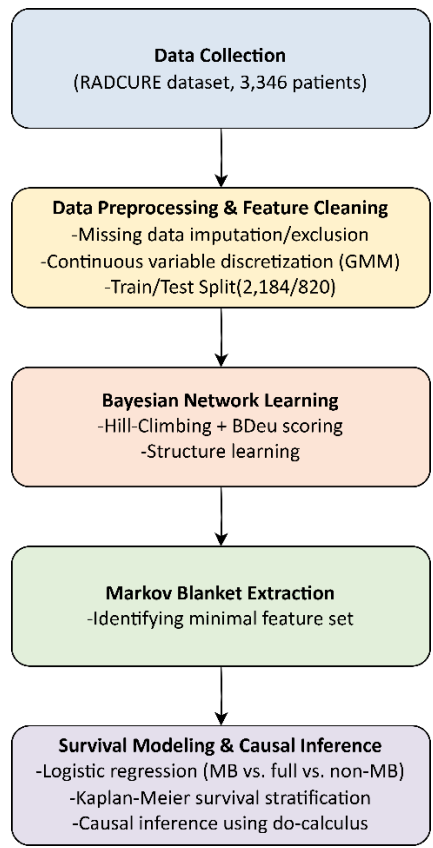

**Figure 1.** Workflow for Bayesian network-based survival modeling in head and neck cancer. The pipeline includes data preprocessing and cleaning, Bayesian network learning with Hill-Climbing and BDeu scoring, extraction of the Markov Blanket (MB) of 2-year survival (SVy2), and downstream logistic regression modeling and causal inference using do-calculus interventions.

## 2.3 Bayesian Network Modeling

We constructed a BN to model conditional dependencies among clinical and anatomical features and to explore the underlying structure contributing to SVy2. Our modeling framework closely follows the structure-learning and preprocessing pipeline described by Chamseddine *et al* (16), adapted here to the H&N cancer context. The BN structure was learned using the Hill-Climbing search algorithm with the Bayesian Dirichlet equivalent uniform (BDeu) scoring metric (10). A maximum in-degree of five was enforced to limit model complexity and improve interpretability. Prior to structure learning, continuous variables were transformed using the Yeo-Johnson power transform (17) and discretized using a Gaussian Mixture Model (GMM)-based approach. The number of bins was selected based on the Akaike Information Criterion (AIC) (18), and the resulting bin edges learned on the training set were applied consistently to the test set. For highly correlated variable pairs (Pearson $|r| > 0.7$), one variable was retained, prioritizing treatment-related or clinically interpretable features, to reduce redundancy. Additionally, variables with low mutual information with respect to the SVy2 outcome (bottom 25th percentile) were excluded unless deemed clinically essential.

The BN parameters were estimated using a Bayesian estimator with the BDeu prior. We extracted the MB of the SVy2 node from the learned structure. The MB represents the smallest subset of variables that renders SVy2 conditionally independent of all other nodes in the graph, thereby capturing all relevant information needed for survival prediction. This subset of variables served as the foundation for subsequent comparative analysis.

## 2.4 Evaluation Metrics

To assess the predictive value of the MB, we evaluated its performance using standard classification metrics and a likelihood-based statistical test. We constructed three logistic regression models: one using all features retained after preprocessing ("all features" model), one using only the MB features ("MB model"), and one using all features excluding the MB ("non-MB model"). This comparison assessed how well the MB captured survival-relevant information relative to larger or non-overlapping feature sets. The area under the receiver operating characteristic curve (AUC) was calculated for each model on both the training and test datasets.

To evaluate model fit, a log-likelihood ratio (LLR) test was performed by comparing the MB model to a null model containing only an intercept. The resulting LLR statistic was evaluated using a chi-squared test, with $p < 0.05$ considered statistically significant. Together, these metrics provide a comprehensive evaluation of how well the MB performs both as a minimal predictive subset and as a statistically meaningful structure derived from the BN.

## 2.5 Overall Risk Stratification and Survival Analysis

To evaluate the MB model's overall risk stratification capability, we applied the MB model to the temporally separated test dataset. Predicted survival probabilities were used to assign patients into high- and low-risk groups using the median probability as a threshold. Kaplan-Meier survival curves were generated to visualize survival differences, and group separation was tested using the log-rank test. The concordance index (C-index) was computed on the test dataset to assess risk discrimination over time.

## 2.6 Subgroup Risk Stratification Analysis

To evaluate the model's performance across clinically relevant subgroups, we performed a post hoc subgroup analysis on the test dataset. Subgroups were retained for analysis if the MB model achieved a higher AUC within that subgroup than in the full test dataset, and if the subgroup contained more than 100 patients. Within each subgroup, patients were stratified into high- and low-risk groups using the median predicted survival probability. Kaplan-Meier curves were used to assess survival separation, and C-indices were calculated to quantify discrimination.

## 2.7 Causal Analysis

To explore potential causal relationships between individual features and two-year survival, we performed post hoc interventional inference using the learned BN and the do-calculus framework. Prior to this analysis, we manually reversed any edges directed from $SVy2$ to its predictors, as survival outcomes cannot biologically influence upstream clinical variables. This adjustment ensured that $SVy2$ appeared only as a child node, maintaining biologically plausible causal directionality. We then focused on variables directly connected to $SVy2$ within its MB, representing its immediate parents in the network. For each of these variables, we simulated interventions by fixing them to specific values using do-operations and estimated the resulting probability of survival ($SVy2 = 1$).

## 3. Results

### 3.1 Feature Selection via Bayesian Network Modeling

The MB of the SVy2 node was extracted from the BN. The MB includes the smallest set of variables that renders SVy2 conditionally independent of all others in the graph. In this analysis, the MB consisted of six variables: ECOG performance status, T-stage, disease site, HPV status, GTVp, and treatment modality. These represent the most directly informative variables for survival within the network structure. Variables such as smoking pack-years, age, pathology, GTV-to-vessel distance, and AJCC stage did not appear in the MB and were excluded from further modeling. A visual depiction of the final BN is provided in Figure 2, where SVy2 is highlighted in red, MB variables are shown in orange, and all remaining nodes are rendered in blue.

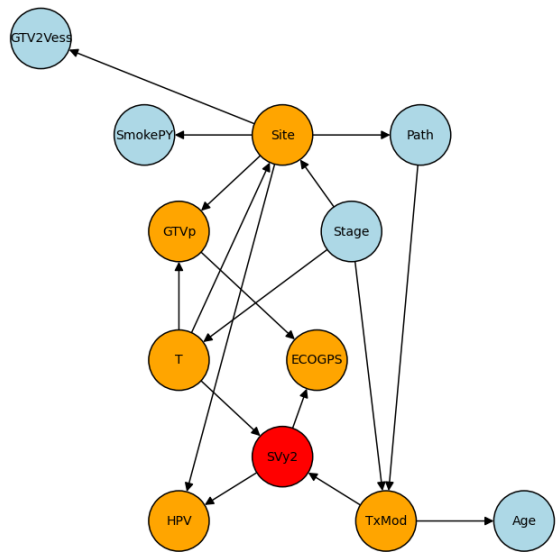

**Figure 2.** Bayesian Network (BN) structure learned from clinical and imaging features for predicting 2-year survival (SVy2) in patients with head and neck cancer. Nodes are color-coded by role: the target variable (SVy2) is shown in red, its Markov Blanket (MB) is shown in orange, and all other variables are shown in blue. Directed edges represent conditional dependencies. Node positions were manually optimized for visual clarity, and the spatial distances between nodes do not convey statistical or temporal relationships.

**Abbreviations:** SVy2, 2-year survival; Site, disease site; GTVp, gross tumor volume; GTV2Vess, abbreviated node label for GTV-to-vessel distance; TxMod, treatment modality; SmokePY, smoking pack-years; ECOGPS, ECOG performance score.

### 3.2 Qualitative Analysis of the Learned BN Structure

To evaluate the interpretability of the learned network, we qualitatively analyzed key edges among variables within the MB and broader graph structure. Table 2 summarizes the interpretation of each edge based on existing clinical knowledge, biological plausibility, or data-driven artifacts. Most connections were found to be clinically established or plausible, supporting the interpretability of the learned structure.

**Table 2.** Interpretation of representative edges in the learned Bayesian Network. Each edge is interpreted based on clinical reasoning or known associations and categorized by evidence type (Clinically Established, Plausible, or Likely Data Artifact). The final column provides contextual justification and relevant supporting references where applicable. These interpretations support the model's clinical coherence and highlight both meaningful and spurious dependencies captured in the learned structure. Edge classification categories were adapted from Chamseddine *et al* (16) to support clinical interpretation of the BN structure.

| Edge | Interpretation | Evidence Type | Comment |
|---|---|---|---|
| Site → Path | Certain anatomical sites have dominant histologies. | Clinically Established | Squamous histology dominates oropharynx/larynx; adenocarcinomas are rare in these sites (19). |
| Site → HPV | HPV-positivity is highly dependent on primary tumor site. | Clinically Established | Oropharynx is the most HPV-associated subsite (8). |
| Site → Stage | Stage distribution differs by site due to anatomical constraints and detection timing. | Plausible | Sites like nasopharynx/hypopharynx often diagnosed late (20, 21). |
| Site → T | Certain sites tend to have larger or smaller primary tumors. | Plausible | Reflects tumor growth behavior, anatomical constraints, and symptom-driven detection delays by site. |
| Site → GTVp | Primary site influences tumor size, reflecting anatomical growth room. | Plausible | Anatomical space in oral cavity vs. larynx can impact measurable volume. |
| Site → SmokePY | Smoking is more prevalent in cancers of specific sites. | Clinically Established | Larynx and hypopharynx are strongly linked to tobacco use (3). |
| Path → TxMod | Pathology type affects treatment choice. | Clinically Established | SCCs may receive cetuximab when cisplatin is contraindicated; non-SCC histologies managed per alternate NCCN pathways (3). |
| Stage → T | T-stage is a direct component of overall stage. | Clinically Established | AJCC TNM staging system (22). |
| Stage → TxMod | Treatment is escalated in higher stage disease. | Clinically Established | ChemoRT is more likely in advanced stages (3). |
| T → GTVp | Larger T-stages imply greater tumor burden. | Clinically Established | T-stage reflects increasing tumor size and invasion depth, correlating with measured GTV (22). |
| T → SVy2 | Survival is worse in advanced T-stage patients. | Clinically Established | Higher T-stage correlates with poorer prognosis due to greater tumor burden (2). |
| GTVp → ECOG PS | Larger tumors may impair function, worsening performance status. | Plausible | Mass effect, dysphagia, or pain can reduce ECOG status; not commonly quantified. |
| SVy2 → ECOGPS | Not a causal link; reversed in do-analysis. | Likely Data Artifact / Correlation | Patients with poorer ECOG performance status tend to have decreased survival; survival cannot causally influence baseline ECOG. |
| SVy2 → HPV | Not causal; reflects improved survival in HPV-positive patients. | Clinically Established (but reversed) | HPV-positive status is a favorable prognostic factor in oropharyngeal cancers (8). |
| TxMod → Age | Older patients may receive less aggressive therapy. | Clinically Established | Age-adjusted treatment selection is standard in H&N oncology (4, 3). |
| TxMod → SVy2 | Treatment type impacts survival but confounded. | Plausible | ChemoRT may improve survival, but retrospective data is subject to selection bias (4). |
| GTV-to-vessel distance → Site | Spatial proximity to vessels varies anatomically by site. | Clinically Established | Hypopharynx, oropharynx are anatomically closer to carotid arteries and other critical vessels (23). |

**Note:** SVy2, 2-year survival; Site, disease site; GTVp, primary gross tumor volume; GTV-to-vessel distance, distance from GTV to nearest vessel; TxMod, treatment modality; SmokePY, smoking pack-years; ECOGPS, ECOG performance score

### 3.3 MB Feature Set Enables Compact Survival Prediction

The all-features model achieved a train AUC of 0.77 and a test AUC of 0.65, suggesting good fit on the training set but limited generalizability. In comparison, the MB model achieved a slightly lower train AUC of 0.71 but similar test AUC of 0.65, indicating reduced overfitting. The non-MB model, which excluded all MB variables, performed worse with a test AUC of 0.58, supporting the MB's value in isolating clinically meaningful predictors. Figure 3 shows the corresponding ROC curves for all three models.

Reclassification metrics further confirmed that the MB model outperformed the null model, with an NRI of 0.16 and an IDI of 0.08 on the test dataset (training: 0.27 and 0.13, respectively). Finally, the LLR values of 206.64 (training) and 70.32 (test), both with $p < 0.01$, validated the MB's statistical significance relative to a null model. Together, these results demonstrate that the MB offers a compact yet predictive representation of survival-relevant variables. These findings demonstrate that the MB achieves near-optimal discrimination with a reduced and clinically interpretable feature set.

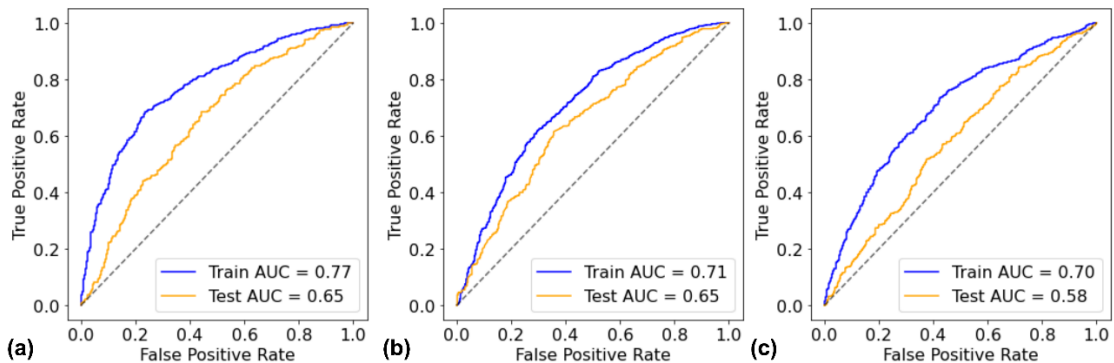

**Figure 3.** Receiver operating characteristic (ROC) curves for logistic regression models predicting 2-year survival (SVy2) using different feature sets. Panel **(a)** shows performance using all features retained after preprocessing; **(b)** shows performance using only features in the Markov Blanket (MB) of SVy2; and **(c)** shows performance using all features excluding the MB. Train and test area under the curve (AUC) values are reported in each panel. The diagonal dashed line represents random classification.

### 3.4 Overall Risk Stratification and Survival Curve Analysis

Using the MB model, patients in the test dataset were stratified into high- and low-risk groups based on the median predicted survival probability. The resulting Kaplan-Meier survival curves, shown in figure 4a, demonstrate clear separation between risk groups. The high-risk group exhibited significantly worse two-year survival compared to the low-risk group (log-rank $p < 0.01$). The C-index for the full test set was 0.78, indicating strong alignment between predicted risk and observed survival times. These findings confirm that the MB-derived model effectively stratifies survival risk across the full cohort.

## 3.5 Subgroup Risk Stratification Analysis

We next evaluated the MB model's performance within key clinical subgroups. Subgroups were selected based on improved AUC compared to the full cohort and a minimum sample size of 100 patients. Three such subgroups were analyzed: HPV-negative patients, patients with T4-stage tumors, and those with GTVp greater the training set median.

As shown in figure 4b–d, the MB model achieved robust risk stratification across all three subgroups. For the HPV-negative subgroup (figure 4b), the C-index was 0.76 and the log-rank p-value < 0.01. In the T4 subgroup (figure 4c), the model achieved a C-index of 0.80 with a log-rank p-value < 0.01. The large-GTV subgroup (figure 4d) yielded a C-index of 0.75 and log-rank $p < 0.01$. In each case, the MB model identified distinct survival trajectories for high- and low-risk groups. ROC curves for these subgroups are shown in supplementary figure S1 a, with corresponding AUCs of 0.69 (HPV-negative), 0.69 (T4), and 0.67 (large GTVp), respectively.

Interestingly, while stratification performance was strong in these higher-risk subgroups, exploratory analyses revealed less pronounced risk separation in more favorable-risk patients. For example, in those with HPV-positive tumors, T1-stage disease, or low GTVp survival curves for high- and low-risk groups were more closely aligned, with C-indices of 0.76, 0.63, and 0.66, respectively. The log-rank test was not statistically significant in the T1 subgroup ($p = 0.22$), further indicating limited discriminatory power in this setting (see Supplementary Figure S2). ROC curves for these subgroups are shown in supplementary figure S1 b. These findings suggest that the MB-derived model is particularly effective at stratifying risk in aggressive disease phenotypes, where outcome heterogeneity is more pronounced.

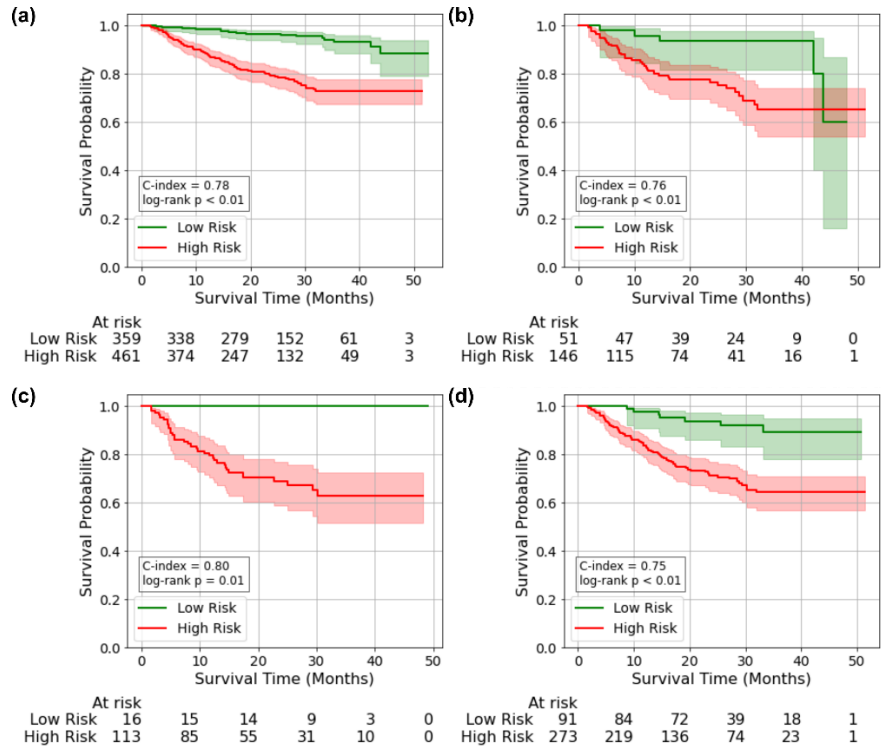

**Figure 4.** Kaplan-Meier survival curves for MB-based risk stratification. Patients were stratified into high- and low-risk groups using the median predicted survival probability from the Markov Blanket (MB) logistic regression model. **(a)** Entire test dataset; **(b)** HPV-negative subgroup; **(c)** T4-stage subgroup; **(d)** Large primary tumor volume subgroup (GTVp > median). In each panel, shaded regions represent 95% confidence intervals. Concordance indices (C-index) and log-rank p-values are reported in the lower left of each plot. At-risk patient counts are shown at regular time intervals. The MB model effectively stratifies survival risk across the full cohort and clinically relevant subgroups.

**3.6 Causal Analysis**

We performed interventional inference on the final BN to estimate the causal impact of four direct MB variables on SVy2: ECOG performance status, T stage, HPV status, and treatment modality. Survival probability declined consistently with worsening ECOG scores, ranging from 0.83 at ECOG 0 to 0.46 at ECOG 4, in line with clinical expectations. HPV-positive patients had a higher probability (0.83) compared to HPV-negative individuals (0.69). Among treatment modalities, concurrent chemoradiation yielded the highest estimated survival probability (0.84), while RT alone (0.72) and EGFR-based regimens (0.67) were associated with lower estimates.

T stage demonstrated a non-monotonic relationship: the highest survival probability was observed at T2 (0.85), with lower values at T3 (0.76) and T4 (0.57). Surprisingly, T0 and T1 stages exhibited lower survival probabilities than T2, despite being considered early stage; this may reflect small sample sizes or residual confounding. Figure 5 illustrates these trends across the four variables, highlighting their estimated causal influence on survival probabilities.

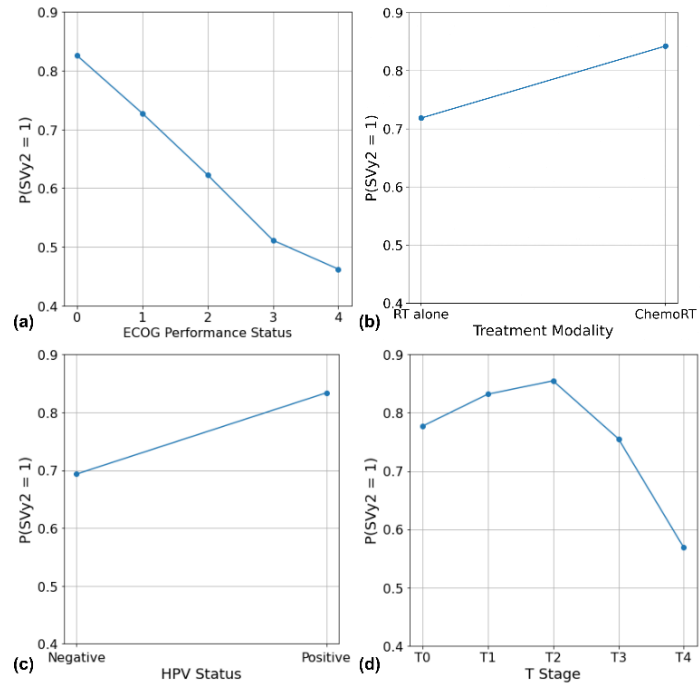

**Figure 5.** Causal effects of variables in the Markov Blanket of 2-year survival (SVy2), estimated using interventional inference with the learned Bayesian Network. Panels show predicted survival probabilities under simulated interventions on **(a)** ECOG performance status, **(b)** treatment modality, **(c)** HPV status, and **(d)** T-stage. Lower ECOG scores and HPV-positivity are associated with higher survival probabilities. Treatment modality and T-stage exhibit non-monotonic trends, suggesting complex interactions worthy of further investigation.

## 4. Discussion

In this study, we developed and validated a BN model to analyze survival outcomes in patients with H&N cancer. The BN effectively captured clinically meaningful dependencies among patient, disease, and treatment variables, while MB of SVy2 emerged as a compact yet predictive subset of features. MB-based logistic regression models offered competitive performance while maintaining interpretability, as demonstrated by strong KM separation and C-indices comparable to deep learning models. The resulting risk scores stratified patients into high- and low-risk groups with significantly different survival trajectories in both the test dataset and in high-risk subgroups. We also applied do-calculus-based causal inference to quantify the effect of individual clinical variables such as ECOG performance status, HPV status, T-stage, and treatment modality on the probability of 2-year survival.

The structure of the learned Bayesian Network provides insights into how patient, disease, and treatment factors interact in H&N cancer. For example, ECOG performance status appears as a child of GTVp, reflecting the idea that larger tumors may negatively impact patient functioning. HPV status is modeled as a child of disease site, reflecting the known association between oropharyngeal tumors and HPV-positivity (8). Treatment modality is represented as a common parent of both age and SVy2, indicating that age-stratified decisions (e.g., omitting chemotherapy in elderly patients) may influence survival, consistent with clinical practice (4, 3). The non-monotonic pattern in T-stage survival likely reflects biological heterogeneity or residual confounding, underscoring the need for caution when interpreting observational causal structures. Additionally, variables such as GTV-to-vessel distance, though not directly associated with survival, were coherently placed within the graph, reinforcing the BN's capacity to incorporate spatial anatomical features in a clinically coherent manner. Together, these relationships support the BN's value as both a predictive tool and a clinical reasoning aid.

Notably, the MB model demonstrated especially strong risk stratification in high-risk clinical subgroups such as T4 tumors, HPV-negative disease, and patients with large GTVp volumes. In contrast, stratification was less effective in lower-risk groups (e.g., T0, HPV-positive, or small-volume disease), where survival curves for high- and low-risk groups overlapped considerably. This pattern suggests that the MB-derived features are particularly informative in aggressive disease settings, where clinical heterogeneity is more pronounced. In contrast, for favorable-risk subgroups, where survival is uniformly high, the model faced a 'ceiling effect', making it harder to differentiate outcomes across risk strata. These findings highlight the potential of the MB model to complement existing risk stratification tools by refining prognostic estimates in challenging clinical contexts.

Survival modeling in H&N cancer has been extensively studied using deep learning and radiomics-based approaches. For example, Kim *et al* applied DeepSurv to oral squamous cell carcinoma and reported a test C-index of 0.78 using 255 patients (24), while Qayyum *et al* used deep PET/CT segmentation and random forest models to achieve a C-index of 0.84 on the HECKTOR21 dataset (25). More recently, Chen *et al* employed a vision transformer-based architecture to predict four survival endpoints in oropharyngeal cancer, achieving C-indices between 0.76 and 0.78, with AUCs up to 0.84 for time-specific outcomes (26). Salmanpour *et al* further explored radiomics fusion methods to predict progression-free survival, though model performance declined on external testing (27). Ma *et al* evaluated a DenseNet-based architecture on the HECKTOR22 dataset, reporting an internal C-index of 0.69 and an external C-index of 0.63, despite attempts at feature fusion and optimization (28). While these models demonstrate strong predictive ability, they rely heavily on high-dimensional imaging, deep feature fusion, and often lack interpretability. In contrast, our MB-based model used only six clinical variables, yet achieved a test C-index of 0.78 overall and up to 0.80 in high-risk subgroups—comparable to or exceeding several of these deep learning benchmarks. Notably, our model outperformed 9 of the top 12 submissions in the RADCURE survival prediction challenge (14) in terms of C-index, despite using no imaging features and a larger test set of 820 patients. These results underscore the value of probabilistic feature selection for building interpretable and clinically robust survival models.

While our model's AUC was modest, this metric reflects a binary framing of survival and does not fully capture the richness of time-to-event data. In contrast, the C-index accounts for concordant risk ranking across censored and uncensored survival times, making it more aligned with clinical utility. This explains the observed divergence between AUC and C-index, and reinforces the need to evaluate survival models using metrics sensitive to event timing, rather than binary thresholds alone.

BNs offer a transparent and interpretable framework for modeling conditional dependencies among clinical and treatment variables. In radiation oncology, BNs have been explored primarily for toxicity prediction, but their application to survival endpoints in H&N cancer has been limited. Luo *et al* demonstrated the value of BNs for modeling radiation pneumonitis in non-small-cell lung cancer (NSCLC), both independently and jointly with tumor local control, capturing biophysical interactions and trade-offs between efficacy and toxicity (29, 30). Similarly, Chamseddine *et al* used BN structure learning to identify key predictors of radiation-induced brain necrosis, showcasing their ability to uncover clinically meaningful associations in high-dimensional datasets (16). Other applications of BNs in radiotherapy include treatment plan error detection (31, 32), response-adaptive planning (33), and recurrence prediction (34). While these efforts reinforce the versatility of BNs in oncology, our study is among the first to apply this framework to survival prediction in H&N cancer and to integrate causal inference for interpreting variable effects within a learned BN structure. Such integration allows for both accurate prediction and mechanistic insights, supporting its potential utility in clinical decision-making and digital twin frameworks. We leveraged a large, well-annotated multi-institutional cohort with a curated set of clinical, anatomical, and treatment-related variables. The use of temporal validation strengthens the generalizability of our findings. Importantly, we adopted a hybrid approach that combines probabilistic graphical modeling (BN) with classical statistical validation (AUC, LLR) and causal inference (do-calculus). By identifying and analyzing the MB of survival, we reduce dimensionality without sacrificing performance, while gaining interpretability. The incorporation of causal graphs further enables simulation of hypothetical interventions, which represents a critical step toward personalized treatment planning. For instance, if a patient is initially predicted to be high-risk, the model can estimate counterfactual survival probabilities under alternative treatment (e.g., switching from RT alone to ChemoRT), enabling exploration of personalized therapeutic options.

Several limitations must be acknowledged. First, our models excluded detailed radiotherapy plan or dose-related features, which are known to affect survival outcomes (35–37). Second, the current study was restricted to a single dataset (albeit temporally split), limiting external validation. Third, the discretization of continuous variables, while necessary for BN learning, may reduce modeling granularity. Fourth, despite using causal graphs to estimate interventional effects, residual confounding from unmeasured variables remains possible. Fifth, patients with less than 24 months of follow-up and no confirmed death were conservatively labeled as SVy2 = 0, which may underestimate survival in censored cases and attenuate classifier performance, particularly in the temporally defined test dataset. Finally, we excluded "Unknown" levels during causal analysis to avoid interpretational artifacts, which may introduce bias if those values are systematically missing.

Future research could explore the integration of advanced, interpretable causal machine learning methods such as causal forests, targeted maximum likelihood estimation (TMLE), and structural causal models (SCMs). These approaches offer the potential to better capture heterogeneous treatment effects and enhance predictive performance while maintaining transparency and clinical interpretability.

This analysis motivates several next steps for expanding both model capabilities and clinical relevance. Incorporating plan-based dosimetric features and imaging-derived biomarkers—such as PET/CT radiomics (38–41) or MRI-derived texture features (42, 43) could further refine survival modeling. Beyond imaging, integrating genomic data is a natural next step, with studies like Jain *et al* highlighting how nonenhancing tumor regions, and epidermal growth factor receptor (EGFR) mutation status in glioblastoma can enrich survival models when paired with clinical and imaging phenotypes (44). Our framework can also be naturally extended to simulate digital twins: patient-specific Bayesian models capable of evaluating counterfactual scenarios. This aligns with recent efforts in brain tumors, where digital twins have been developed to personalize radiotherapy under uncertainty (45). Ultimately, we envision BNs serving as interpretable backbone models for clinical decision support—integrating structured domain knowledge with data-driven learning to enable transparent, individualized, and evidence-based guidance in radiation oncology.

## 5. Conclusion

In this study, we present a BN framework for survival prediction in H&N cancer, integrating probabilistic graphical modeling, classical statistical validation, and causal inference. Although 2-year survival was the primary endpoint, the model structure is readily adaptable to other clinically relevant outcomes. Our approach captures clinically meaningful dependencies and enables both accurate prediction and mechanistic interpretation of survival outcomes. The compact MB enables dimensionality reduction without compromising performance, while do-calculus supports individualized counterfactual reasoning—laying the foundation for future digital twin applications. Moving forward, we aim to expand this framework by incorporating dosimetric and imaging-derived features and exploring advanced causal machine learning methods to further enhance interpretability and clinical impact.

# References

1. Sung H, Ferlay J, Siegel RL, *et al.* Global Cancer Statistics 2020: GLOBOCAN Estimates of Incidence and Mortality Worldwide for 36 Cancers in 185 Countries. *CA A Cancer J Clinicians*. 2021;71:209–249.

2. Lydiatt WM, Patel SG, O'Sullivan B, *et al.* Head and neck cancers—major changes in the American Joint Committee on cancer eighth edition cancer staging manual. *CA A Cancer J Clinicians*. 2017;67:122–137.

3. Colevas AD, Cmelak AJ, Pfister DG, *et al.* NCCN Guidelines® Insights: Head and Neck Cancers, Version 2.2025: Featured Updates to the NCCN Guidelines. *Journal of the National Comprehensive Cancer Network*. 2025;23:2–11.

4. Pignon J-P, Maître AL, Maillard E, *et al.* Meta-analysis of chemotherapy in head and neck cancer (MACH-NC): An update on 93 randomised trials and 17,346 patients. *Radiotherapy and Oncology*. 2009;92:4–14.

5. Nguyen-Tan PF, Zhang Q, Ang KK, *et al.* Randomized Phase III Trial to Test Accelerated Versus Standard Fractionation in Combination With Concurrent Cisplatin for Head and Neck Carcinomas in the Radiation Therapy Oncology Group 0129 Trial: Long-Term Report of Efficacy and Toxicity. *JCO*. 2014;32:3858–3867.

6. Muzaffar J, Bari S, Kirtane K, *et al.* Recent Advances and Future Directions in Clinical Management of Head and Neck Squamous Cell Carcinoma. *Cancers*. 2021;13:338.

7. Zhong N-N, Wang H-Q, Huang X-Y, *et al.* Enhancing head and neck tumor management with artificial intelligence: Integration and perspectives. *Seminars in Cancer Biology*. 2023;95:52–74.

8. Ang KK, Harris J, Wheeler R, *et al.* Human Papillomavirus and Survival of Patients with Oropharyngeal Cancer. *N Engl J Med*. 2010;363:24–35.

9. Pearl J. *Probabilistic Reasoning in Intelligent Systems*. Morgan Kaufmann; 1988.

10. Koller D, Friedman N. *Probabilistic graphical models: principles and techniques*. Nachdr. Cambridge, Mass.: MIT Press; 2010.

11. Pearl J. *Probabilistic Reasoning in Intelligent Systems: Networks of Plausible Inference*. 1. Aufl. s.l.: Elsevier Reference Monographs; 2014.

12. Tsamardinos I, Aliferis CF. Algorithms for large scale Markov blanket discovery. In: ; 2003.

13. Welch ML, Kim S, Hope AJ, *et al.* RADCURE: An open-source head and neck cancer CT dataset for clinical radiation therapy insights. *Medical Physics*. 2024;51:3101–3109.

14. Kazmierski M, Welch M, Kim S, *et al.* Multi-institutional Prognostic Modeling in Head and Neck Cancer: Evaluating Impact and Generalizability of Deep Learning and Radiomics. *Cancer Research Communications*. 2023;3:1140–1151.

15. Wasserthal J, Breit H-C, Meyer MT, *et al.* TotalSegmentator: Robust Segmentation of 104 Anatomic Structures in CT Images. *Radiology: Artificial Intelligence*. 2023;5:e230024.

16. Chamseddine I, Shah K, Lee H, *et al.* Decoding Patient Heterogeneity Influencing Radiation-Induced Brain Necrosis. *Clinical Cancer Research*. 2024;30:4424–4433.

17. Yeo I-K, Johnson RA. A new family of power transformations to improve normality or symmetry. *Biometrika*. 2000;87:954–959.

18. Akaike H. A new look at the statistical model identification. *IEEE Trans. Automat. Contr.* 1974;19:716–723.

19. López F, Williams MD, Cardesa A, *et al.* How phenotype guides management of non-conventional squamous cell carcinomas of the larynx? *Eur Arch Otorhinolaryngol*. 2017;274:2709–2726.

20. Razak ARA, Siu LL, Liu F-F, *et al.* Nasopharyngeal carcinoma: The next challenges. *European Journal of Cancer*. 2010;46:1967–1978.

21. Takes RP, Strojan P, Silver CE, *et al.* Current trends in initial management of hypopharyngeal cancer: The declining use of open surgery. *Head & Neck*. 2012;34:270–281.

22. American Joint Committee on Cancer. *AJCC cancer staging manual*. Eighth edition, corrected at 3rd printing. (Amin MB, Greene FL, Edge SB, eds.). Chicago, IL: AJCC, American Joint Committee on Cancer; 2017.

23. Grégoire V, Ang K, Budach W, *et al.* Delineation of the neck node levels for head and neck tumors: A 2013 update. DAHANCA, EORTC, HKNPCSG, NCIC CTG, NCRI, RTOG, TROG consensus guidelines. *Radiotherapy and Oncology*. 2014;110:172–181.

24. Kim DW, Lee S, Kwon S, *et al.* Deep learning-based survival prediction of oral cancer patients. *Sci Rep*. 2019;9:6994.

25. Qayyum A, Benzinou A, Razzak I, *et al.* 3D-IncNet: Head and Neck (H&N) Primary Tumors Segmentation and Survival Prediction. *IEEE J. Biomed. Health Inform.* 2024;28:1185–1194.

26. Chen M, Wang K, Wang J. Vision Transformer-Based Multilabel Survival Prediction for Oropharynx Cancer After Radiation Therapy. *International Journal of Radiation Oncology*Biology*Physics*. 2024;118:1123–1134.

27. Salmanpour MR, Hosseinzadeh M, Rezaeijo SM, *et al.* Fusion-based tensor radiomics using reproducible features: Application to survival prediction in head and neck cancer. *Computer Methods and Programs in Biomedicine*. 2023;240:107714.

28. Ma B, Guo J, Dijk LVV, *et al.* PET and CT based DenseNet outperforms advanced deep learning models for outcome prediction of oropharyngeal cancer. *Radiotherapy and Oncology*. 2025;207:110852.

29. Luo Y, El Naqa I, McShan DL, *et al.* Unraveling biophysical interactions of radiation pneumonitis in non-small-cell lung cancer via Bayesian network analysis. *Radiotherapy and Oncology*. 2017;123:85–92.

30. Luo Y, McShan DL, Matuszak MM, *et al.* A multiobjective Bayesian networks approach for joint prediction of tumor local control and radiation pneumonitis in nonsmall-cell lung cancer ( NSCLC ) for response-adapted radiotherapy. *Medical Physics*. 2018;45:3980–3995.

31. Kalet AM, Gennari JH, Ford EC, *et al.* Bayesian network models for error detection in radiotherapy plans. *Phys. Med. Biol.* 2015;60:2735–2749.

32. Luk SMH, Meyer J, Young LA, *et al.* Characterization of a Bayesian network-based radiotherapy plan verification model. *Medical Physics*. 2019;46:2006–2014.

33. Ajdari A, Liao Z, Mohan R, *et al.* Personalized mid-course FDG-PET based adaptive treatment planning for non-small cell lung cancer using machine learning and optimization. *Phys. Med. Biol.* 2022;67:185015.

34. Osong B, Masciocchi C, Damiani A, *et al.* Bayesian network structure for predicting local tumor recurrence in rectal cancer patients treated with neoadjuvant chemoradiation followed by surgery. *Physics and Imaging in Radiation Oncology*. 2022;22:1–7.

35. Shaikh T, Handorf EA, Murphy CT, *et al.* The Impact of Radiation Treatment Time on Survival in Patients With Head and Neck Cancer. *International Journal of Radiation Oncology\*Biology\*Physics*. 2016;96:967–975.

36. Rosenthal DI, Mohamed ASR, Garden AS, *et al.* Final Report of a Prospective Randomized Trial to Evaluate the Dose-Response Relationship for Postoperative Radiation Therapy and Pathologic Risk Groups in Patients With Head and Neck Cancer. *International Journal of Radiation Oncology\*Biology\*Physics*. 2017;98:1002–1011.

37. Mazul AL, Stepan KO, Barrett TF, *et al.* Duration of radiation therapy is associated with worse survival in head and neck cancer. *Oral Oncology*. 2020;108:104819.

38. Liu Z, Cao Y, Diao W, *et al.* Radiomics-based prediction of survival in patients with head and neck squamous cell carcinoma based on pre- and post-treatment 18F-PET/CT. *Aging*. 2020;12:14593–14619.

39. Naser MA, Wahid KA, Mohamed ASR, *et al.* Progression Free Survival Prediction for Head and Neck Cancer Using Deep Learning Based on Clinical and PET/CT Imaging Data. In: Andrearczyk V, Oreiller V, Hatt M, *et al.*, eds. *Head and Neck Tumor Segmentation and Outcome Prediction*.Vol 13209. Lecture Notes in Computer Science. Cham: Springer International Publishing; 2022:287–299.

40. Wang Y, Lombardo E, Avanzo M, *et al.* Deep learning based time-to-event analysis with PET, CT and joint PET/CT for head and neck cancer prognosis. *Computer Methods and Programs in Biomedicine*. 2022;222:106948.

41. Salahuddin Z, Chen Y, Zhong X, *et al.* From Head and Neck Tumour and Lymph Node Segmentation to Survival Prediction on PET/CT: An End-to-End Framework Featuring Uncertainty, Fairness, and Multi-Region Multi-Modal Radiomics. *Cancers*. 2023;15:1932.

42. Chawla S, Kim S, Loevner LA, *et al.* Prediction of Disease-Free Survival in Patients with Squamous Cell Carcinomas of the Head and Neck Using Dynamic Contrast-Enhanced MR Imaging. *AJNR Am J Neuroradiol*. 2011;32:778–784.

43. Bae S, Choi YS, Ahn SS, *et al.* Radiomic MRI Phenotyping of Glioblastoma: Improving Survival Prediction. *Radiology*. 2018;289:797–806.

44. Jain R, Poisson LM, Gutman D, *et al.* Outcome Prediction in Patients with Glioblastoma by Using Imaging, Clinical, and Genomic Biomarkers: Focus on the Nonenhancing Component of the Tumor. *Radiology*. 2014;272:484–493.

45. Chaudhuri A, Pash G, Hormuth DA, *et al.* Predictive digital twin for optimizing patient-specific radiotherapy regimens under uncertainty in high-grade gliomas. *Front. Artif. Intell.* 2023;6:1222612.

# Supplementary Materials

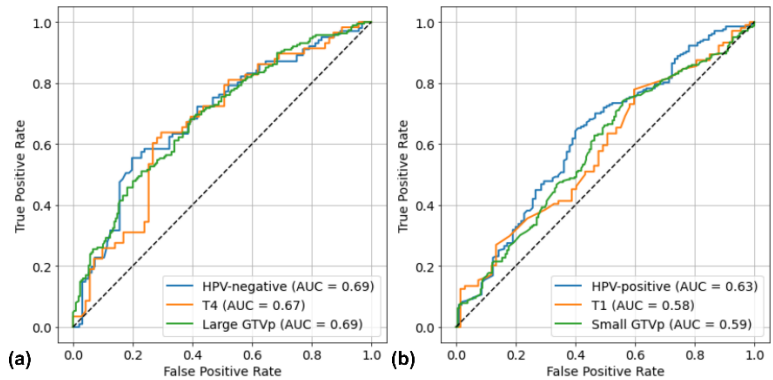

**Figure S1.** ROC curves for MB model predictions in high- and low-risk subgroups. **(a)** Subgroups where the MB model achieved higher AUCs, **(b)** Subgroups with lower AUCs.

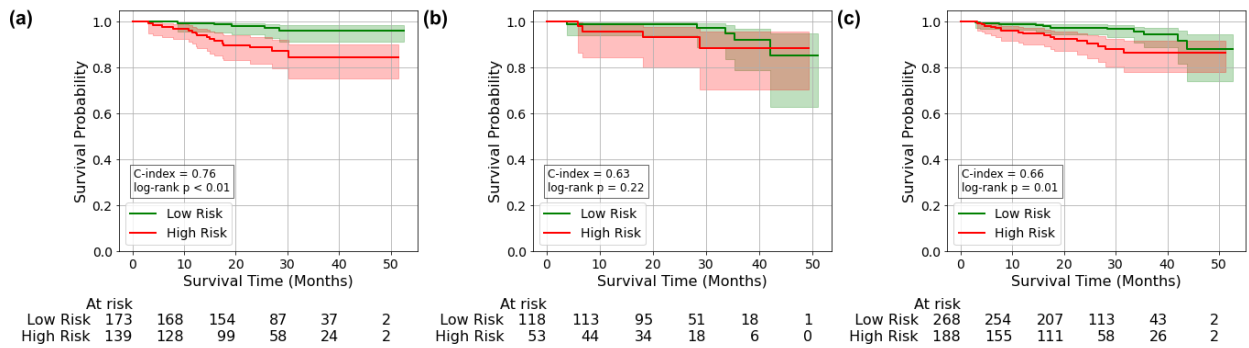

**Figure S2.** Kaplan-Meier survival curves for MB-based risk stratification in lower-risk subgroups. **(a)** HPV-positive patients, **(b)** T1-stage tumors, and **(c)** small-volume tumors (GTVp ≤ median). Risk groups were defined by the median predicted probability of 2-year survival. C-indices and log-rank p-values are reported within each panel. Compared to high-risk subgroups, survival stratification is less pronounced in these more favorable patient populations.